\documentstyle[psfig]{mn-1.4}

\title[Broad-band study of the red quasar SAXJ~1353.9$+$1820]
{Optical, near-infrared and hard X--ray observations of SAXJ~1353.9$+$1820: a red quasar
\thanks{Based on observations performed with 
the Italian National Telescope, which is operated on the island of La Palma by the 
Centro Galileo Galilei/CNAA at the Spanish Observatorio del Roque de Los 
Muchachos of the Instituto de Astrofisica de Canarias
}}

\author[C. Vignali et al.]{
C. Vignali$^{1,2}$, 
M. Mignoli$^{2}$, 
A. Comastri$^{2}$, 
R. Maiolino$^{3}$, 
F. Fiore$^{4,5,6}$ \\ 
$^1$ Dipartimento di Astronomia, Universit\`a di Bologna, 
Via Ranzani 1, I--40127 Bologna, Italy \\
$^2$ Osservatorio Astronomico di Bologna, 
Via Ranzani 1, I--40127 Bologna, Italy \\
$^3$ Osservatorio Astrofisico di Arcetri, 
Largo E. Fermi 5, I--50125 Firenze, Italy \\
$^4$ Osservatorio Astronomico di Roma, 
Via Frascati 33, I--00044 Monteporzio, Italy \\
$^5$ BeppoSAX Science Data Center, 
Via Corcolle 19, I--00131 Roma, Italy \\
$^6$ Harvard-Smithsonian Center of Astrophysics, 
60 Garden Street, Cambridge MA 02138 USA \\
}
\date{Accepted ...
      Received ...;
      in original form ...}

\pagerange{\pageref{firstpage}--\pageref{lastpage}}
\pubyear{1999}

\begin{document}

\maketitle

\label{firstpage}

\begin{abstract}

We present the results of a follow--up {\it ASCA} observation 
and multicolour optical and near-infrared photometry carried out
at the 3.5-m Italian National Telescope Galileo of SAXJ~1353.9$+$1820.
This object, serendipitously discovered by {\it BeppoSAX}
in the 5--10 keV band, has been spectroscopically identified as a 
red quasar at $z$=0.217.
The combined X--ray and optical--infrared data reveal the 
presence of a moderately luminous X--ray source ($\sim$ 10$^{44}$ erg s$^{-1}$)
obscured by a column density of the order of 10$^{22}$ cm$^{-2}$ 
in a otherwise optically passive early-type galaxy. 
The implications for the nature of red quasars and their possible 
contribution to the hard X--ray background are briefly outlined.

\end{abstract}


\section{Introduction}

The {\it BeppoSAX} High Energy Large Area Survey (HELLAS) has discovered a large 
population of hard X--ray sources, 
which account for about 20--30 per cent of the cosmic X--ray background (XRB) in 
the 5--10 keV energy range (Fiore et al. 1999). 
This band provides an efficient tool to discriminate between accretion-powered sources, 
like the Active Galactic Nuclei (AGN) are generally thought to be, and sources dominated by the starlight 
component. Moreover, hard X--ray selection 
is less affected by obscuration with respect to other bands, making 
possible the study of the nuclear 
continuum without any relevant contamination by reprocessed radiation. 
Indeed the optical identification process indicates that the large majority
of the HELLAS sources are AGN (La Franca et al., in preparation). 
Among these, SAXJ~1353.9$+$1820 can be considered one of the most 
intriguing sources. 
After its discovery in the context of the HELLAS survey, it was 
spectroscopically identified as a radio-quiet AGN at a redshift of 0.2166 
(Fiore et al. 1999).  Its optical spectrum 
is dominated by starlight, as the presence of H and K plus
Mg~I absorption lines clearly indicates. 
The H$\alpha$ equivalent width (about 90 $\AA$),
the absence of the H$\beta$ and the {\it BeppoSAX} hardness ratio  
are suggestive of significant reddening of the nuclear radiation.
The hard X--ray luminosity, L$_{5-10~keV}$ $\sim$ 1.4 $\times$ 10$^{44}$ 
erg s$^{-1}$, the presence of a broad Balmer line and the optical 
spectrum properties allow to classify this object as a fairly low-luminosity, red quasar. \\
This class of objects, originally discovered in radio--selected samples 
(Smith \& Spinrad 1980), is characterized by red optical colours 
($B-K$ up to 8, Webster et al. 1995).
On the basis of {\it ROSAT} PSPC observations and optical
spectroscopy Kim \& Elvis (1999) pointed out that a significant fraction 
(from a few percent up to 20 per cent) 
of soft X--ray selected radio--quiet quasars belong to this class.
The origin of the observed `redness' may be ascribed to dust absorption, 
to intrinsic red colours ($\ga$ 70 per cent of the red sources of the Parkes sample
has $\la$ 30 per cent of contribution to B--K by their host galaxies, 
Masci et al. 1998) or to an excess of light in K band rather than 
a dust-induced deficit in B (Benn et al. 1998).
Whatever is the origin of the red colours, it is likely  
that a large fraction of quasars could have been missed by the usual 
selection techniques in the optical band (Webster et al. 1995).
If this is the case, red quasars could constitute a sizeable fraction 
of the absorbed AGN population needed to explain the hard X--ray 
background spectrum (i.e. Comastri et al. 1995) especially 
if the optical reddening is associated with X--ray absorption.  
Hard X--ray observations provide the most efficient way to 
select these objects;
indeed already two candidates have been found among the first optical
identifications of HELLAS sources (Fiore et al. 1999).   
In order to better understand the spectral properties of these objects 
we have started a program of multiwavelength follow--up 
observations of a sub--sample of HELLAS sources.
Here we present the first results obtained 
in the X--ray band with {\it ASCA} and at optical--infrared 
wavelengths with the 3.5-m Italian National Telescope Galileo (TNG) at La Palma 
(OIG and ARNICA photometric cameras). 
Throughout the paper $H_{0}$ = 50 km s$^{-1}$ Mpc$^{-1}$ and $q{_0}$ = 0 are assumed.

\section{ASCA data reduction and spectral analysis}
SAXJ~1353.9$+$1820 (RA: 13$^{h}$ 53$\arcmin$ 54$\arcsec$.4, DEC: 18$^{\circ}$ 20$\arcmin$ 16$\arcsec$) 
was observed with the {\it ASCA} satellite (Tanaka, Inoue \& Holt 1994) in January 1999 
for about 60 ks. 
The observation was performed in FAINT mode and then corrected for dark frame error and echo uncertainties 
(Otani \& Dotani 1994).  
The data were screened with the {\sc XSELECT} package (version 1.4b) with standard criteria. 
Spectral analysis on the resulting cleaned data was 
performed with {\sc XSPEC} version 10 (Arnaud et al. 1996). \\
The background-subtracted count rates are 6.0$\pm{0.4}$ $\times$ 10$^{-3}$ counts s$^{-1}$ for 
SIS (Solid-State Spectrometers, Gendreau 1995) 
and 7.4$\pm{0.4}$ $\times$ 10$^{-3}$ counts s$^{-1}$ for GIS (Gas-Scintillation Spectrometers, Makishima et. al 1996). 
Both SIS and GIS spectra were grouped with almost 20 photons  
for each spectral bin in order to apply $\chi^{2}$ statistics. 
Calibration uncertainties in the soft X--ray band have been avoided by selecting only data 
at energies higher than 0.9 keV. No discrepancies have been found between SIS and GIS 
spectral analysis, therefore all the data have been fitted together allowing the relative normalizations 
to be free of varying. 
The uncertainties introduced by background subtraction have been 
carefully checked using both local and blank--sky background spectra
and also varying their normalizations by $\pm$ 10 per cent.
The lack of significant variations for the source count rate 
and spectral shape makes us confident on the robustness of
the results. \\
A simple power law model plus Galactic absorption ($N_{\rm H}$ $\simeq$ 2.05 $\times$ 10$^{20}$ cm$^{-2}$, Dickey \& 
Lockman 1990) leaves some residuals in the fit ($\chi^{2}$=174/158) and
 gives a very flat slope ($\Gamma$ $<$ 0.9).
The addition of an extra cold absorber at the 
redshift of the source (model {\bf (a)} in Table~1) improves the fit 
and the continuum X--ray spectral slope 
is now $\Gamma$ = 1.28$^{+0.23}_{-0.28}$ (errors are at 90 per cent for one interesting parameter, 
or $\Delta\chi^{2}$=2.71, Avni 1976), 
attenuated by a column density $N_{\rm H}$ = 6.14$^{+2.10}_{-4.56}$ $\times$ 10$^{21}$ cm$^{-2}$, assuming cosmic abundances 
(Anders \& Grevesse 1989) and cross sections derived by Balucinska-Church 
\& McCammon (1992). 
The best-fitting spectrum and the confidence contours for 
the absorbed power law model are presented in Fig.~1 and Fig.~2, respectively. 
The unabsorbed 2--10 keV flux and luminosity are
$\sim$ 6.2 $\times$ 10$^{-13}$ erg cm$^{-2}$ s$^{-1}$
and $\sim$ 1.3 $\times$ 10$^{44}$ erg s$^{-1}$. 
The 5--10 keV {\it ASCA} flux is about 40 per cent lower than in {\it BeppoSAX}. 
X--ray variability and/or cross--calibration uncertainties could provide
a likely explanation. \\

\begin{figure}
\psfig{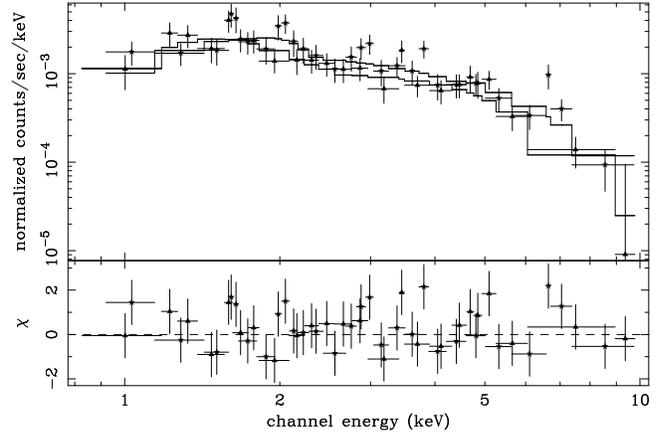}
\caption[]{ASCA SIS $+$ GIS spectrum and relative data/model ratio for
the absorbed power law model (model {\bf (a)} in Table~1).}
\label{fig1}
\end{figure}

\begin{table}
\caption[]{ASCA SIS$+$GIS spectral fits (0.9--10 keV energy range)}
\begin{tabular}{lcccc}
\noalign{\smallskip}
\hline
\noalign{\smallskip}
{\bf Model} & $\Gamma$ & $N_{\rm H}$ & CvrFract & $\chi^{2}$/dof \\
\noalign{\smallskip}
 & & (10$^{21}$ cm$^{-2}$) & (\%) & \\
\noalign{\smallskip}
\hline \hline
\noalign{\smallskip}
{\bf (a)} & 1.28$^{+0.23}_{-0.28}$ & 6.14$^{+2.10}_{-4.56}$ & \dots & 168/157 \\

{\bf (b)} & 1.9 (frozen) & 15.4$^{+3.70}_{-3.20}$ & \dots & 181/158 \\

{\bf (c)} & 1.28$^{+0.43}_{-0.30}$ & 8.55$^{+24.0}_{-6.89}$ & 81$^{+19}_{-44}$ & 168/156 \\

{\bf (d)} & 1.9 (frozen) & 28.7$^{+17.2}_{-5.90}$ & 80$^{+11}_{-8}$ & 171/157 \\

\noalign{\smallskip}
\hline
\end{tabular}
\end{table}

\par\noindent
SAXJ~1353.9$+$1820 does not show any particular feature or any other indication of 
reprocessed radiation. 
Neither the iron K$\alpha$ emission line (the 90 per cent upper limit on the equivalent width being 
330 eV) nor the reflection component (which is basically unconstrained 
by the present data)
do improve the fit.
The  $N_{\rm H}$ value  
(which has been fixed at the quasar redshift but which could lie along the line 
of sight to the QSO) 
is in agreement with the value found by the hardness 
ratio analysis of {\it BeppoSAX} data.

\begin{figure}
\psfig{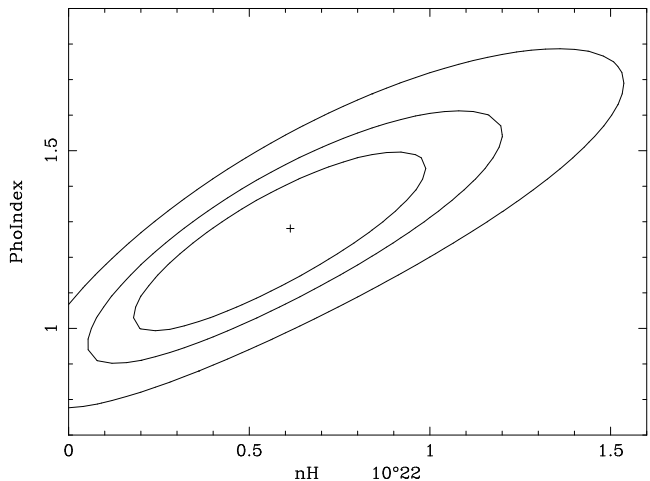}
\caption[]{ASCA confidence contours in the $\Gamma$ -- $N_{{\rm H}_{\rm int}}$ space parameters.}
\label{fig2}
\end{figure}

The best-fitting slope is extremely hard and significantly flatter than the
average slope of Seyfert Galaxies and quasars with similar luminosities
and redshifts (Nandra \& Pounds 1994; Reeves et al. 1997; George et al. 1999). 
Indeed assuming a `canonical' $\Gamma$ = 1.9 value
we were not able to obtain a good fit as relatively large residuals 
are present at high energies (model {\bf (b)}). 
The underlying continuum spectrum could be either intrinsically flat
or flattened by a complex (multicolumn and/or leaky) absorber (Hayashi et al. 1996; Vignali et al. 1998).  
To test this last hypothesis we have fitted a partial covering model (models {\bf (c)} and {\bf (d)} in Table~1), 
where part of the direct component escapes without being absorbed.
While the best-fitting model still requires a flat slope, a steeper continuum
partially absorbed by a column density of $\sim$ 
3 $\times$ 10$^{22}$ cm$^{-2}$ does provide a good fit to the observed 
spectrum.\\

\section{Optical and NIR photometry}
SAXJ~1353.9+1820 has been observed at the 3.5-m National Telescope Galileo 
with the Optical Imager (OIG) during the night of 1999 June 18. 
We carried out optical broad band imaging in the Johnson--Kron--Cousins 
{\it U,~B,~V,~R} and~{\it I} filters. The 
exposure times were respectively 900,~480,~180,~120 and~120 seconds while 
the seeing ranges from 0.9~arcsec (FWHM) in the reddest band to 1.3 arcsec in 
the ultraviolet, with a steady increase through the optical bands giving 
evidence that the blurring is mainly due to atmospheric causes. During the same 
night we observed the standard fields PG~1323$-$086, PG~1633$+$099 and 
SA~110 in order to obtain accurate photometric calibrations and to determine 
the colour terms of the relatively new OIG system. 
All the frames were acquired at airmass $\leq 1.5$. The data 
reduction and analysis has been performed in a standard way using 
{\sc IRAF}\footnote{{\sc IRAF} is distributed by the National Optical Astronomy 
Observatories, which are operated by the Associated Universities for Research 
in Astronomy, Inc. under cooperative agreement with the National Science 
Foundation}
routines. Bias exposures taken at the beginning and at the end of the night were 
stacked, checked for consistency with the overscan region of the scientific 
images and subtracted out. The bias--subtracted frames were then flat--fielded 
using sky flats. The cosmic rays of the CCD region around the target have 
been interactively identified and removed by fitting of the neighbouring pixels. 

\noindent
The photometry has been performed using {\tt apphot}, the Aperture Photometry 
Package available in {\sc IRAF}. The object is clearly extended, and we used 
a quite large aperture radius ($\sim$ 8~arcsec) for all the bands, 
corresponding to a projected distance of about 15 $h_{50}^{-1}$~kpc at the redshift 
of SAXJ~1353.9+1820. Measurements in the J and K-short bands were also made with 
the ARNICA instrument at the same telescope within the framework of a wider 
near--IR follow-up of the HELLAS sources. The data reduction and analysis 
of the near--IR 
data will be discussed in detail in Maiolino et al. (in preparation), here we simply report the resulting 
photometry. 
The results of the combined aperture photometry are presented in Table~2. 

\begin{figure}
\psfig{figure=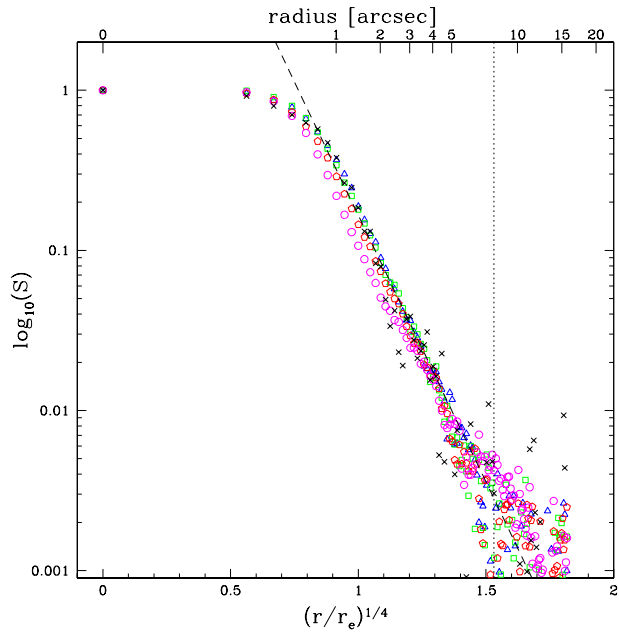,width=0.48\textwidth,angle=0}
\caption[]{The optical surface brightness profiles 
fitted with a de Vaucouleurs r$^{1/4}$ law (dashed line). 
Different symbols are referred to different filters.
The vertical dotted line represents the adopted aperture radius.}
\label{fig3}
\end{figure}

The surface brightness (SB) profiles in the U, B, V, R and 
I filters have been estimated computing a curve of growth for each 
passband with increasing circular apertures (Fig.~3).
An effective radius of $\sim$~1.5 arcsec 
was indipendently derived from all but one fits (for the I--band we obtained a 
slightly more concentrated profile). 
The dashed line in figure represents the $r^{1/4}$ law, which fits 
very well the observed profiles outside of the seeing-dominated region
down to the faintest flux levels.  
The SB profiles are typical for elliptical galaxies suggesting 
that at optical wavelengths there is no evidence of
an unresolved nucleus in SAXJ~1353.9$+$1820.

\begin{table}
\caption{Optical and NIR photometry}
\label{phot}
\begin{center}
\begin{tabular}{lccccccc}
\hline
Filter & {\bf U} & {\bf B} & {\bf V} & {\bf R} & {\bf I} & {\bf J} & {\bf Ks} \\

Mag. & 19.98 & 19.55 & 18.15 & 17.32 & 16.62 & 15.29 & 13.91 \\

Errors & 0.07 & 0.03 & 0.03 & 0.02 & 0.04 & 0.05 & 0.07 \\
\hline
\end{tabular}
\end{center}
\end{table}

\section{Discussion and Conclusions}

The {\it ASCA} observation confirms 
the presence of a bright and moderately 
absorbed ($N_{\rm H}$ $\sim$ 6 $\times$ 10$^{21}$ cm$^{-2}$) 
nucleus with a flat hard X--ray spectrum, in agreement with the BeppoSAX hardness ratio analysis. 
Assuming for the underlying continuum slope the average value 
of quasars in the same energy range, the absorption column density
could be as high as $N_{\rm H}$ $\sim$ 3 $\times$ 10$^{22}$ cm$^{-2}$
if some 20 per cent of the nuclear radiation is not absorbed.
Even if the X--ray data alone does not allow to distinguish 
between the two possibilities, we can safely conclude 
that SAXJ~1353.9$+$1820 harbours a mildly obscured, 
luminous (L$_{\rm 2-10 keV}$ $\simeq$ 1.3 $\times$ 10$^{44}$ erg s$^{-1}$)
active nucleus. 

Not surprisingly high energy observations of quasars 
characterized by similar dust reddened optical continua 
do reveal the presence of absorption by cold 
(IRAS 23060+0505, Brandt et al. 1997a) 
and/or warm (IRAS 13349+2438, Brandt et al. 1997b) gas. 

A basic step forward in understanding the nature of this red quasar
and of red quasars as a whole, is provided  
by optical plus near--IR studies 
(see Maiolino et al., in preparation, for further details).
The surface brightness profiles are consistent with those of an elliptical galaxy, and 
the optical colours (U$-$B=0.43, B$-$V=1.40, V$-$R=0.83 and R$-$I=0.70) 
agree with the properties of a early-type galaxy at $z$=0.2 (Fukugita et al. 1995).

In order to obtain a self--consistent description of the optical-IR properties 
we fitted the photometric points with a two-components model consisting 
of an old stellar population template (10$^{10}$ yr, Bruzual \& Charlot 1993) 
and a moderately absorbed ($A_{\rm V} \simeq$ 2 mag, corresponding
to $N_{\rm H} \simeq$ 4 $\times$ 10$^{21}$ cm$^{-2}$ for Galactic 
dust--to--gas ratio, {\rm i.e.} Bohlin, Savage \& Drake 1978) 
quasar spectrum template (Elvis et al. 1994; Francis et al. 1991). 
As shown in Fig.~4, 
the combination of these two spectra provides a good description of 
the observed data 
(possibly with the exception of the J photometry, which deviates by 1.4 sigma). 
The quality of the fit is acceptable ($\chi^2_{\rm r}$ = 1.3 when 
both the uncertainties in the photometric data and in the template spectra
are taken into account) indicating that most 
of the optical and near--IR flux is dominated by star--light. 

\begin{figure}
\psfig{figure=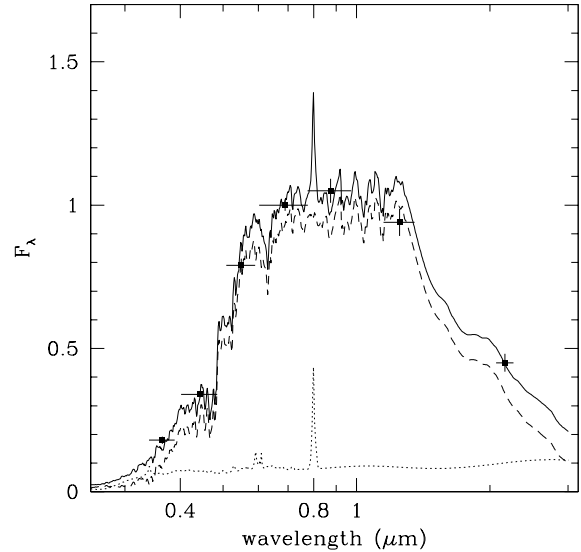,width=0.48\textwidth,angle=0}
\caption[]{SAXJ~1353.9$+$1820 photometric (optical $+$ NIR) points 
fitted with a synthetic model of an evolved early-type galaxy 
(dashed line) plus the contribution of a moderately absorbed 
quasar (dotted line). The sum is represented by the solid line.}
\label{fig4}
\end{figure}

The present results add further evidence on the hypothesis of 
substantial gas and dust absorption as an explanation of the 
observed properties of this red quasar.
Even more interesting is that the active nucleus peers only
at X--ray energies and possibly at wavelenghts longward of 2 $\mu$m (see Fig.~4), 
while at optical wavelengths 
SAXJ~1353.9$+$1820 looks like a normal evolved elliptical galaxy.
If this behaviour applies also to other objects it may well be 
that a significant fraction of obscured AGNs resides
in otherwise normal passive galaxies.
These nuclei would have been completely missed in optical quasar surveys
because of their extended morphology and galaxy-like colours. 
If the column density is of the order of  10$^{22}$ cm$^{-2}$ or higher, 
their fraction could be understimated also in soft X--ray surveys. 
If this is the case, the fraction of red objects 
among radio quiet quasars ($\sim$ 3--20 per cent, Kim \& Elvis 1999) 
should be considered as a lower limit. 
This may have strong implications 
for the XRB synthesis models, which in their 
simplest version (i.e. Madau, Ghisellini \& Fabian 1994; Comastri et al. 1995)
predict a large number of high-luminosity absorbed quasars (called type 2 QSO).
Despite intensive optical searches, these objects appears to be elusive 
indicating a much lower space density than that of 
lower luminosity Seyfert 2 galaxies (Halpern, Eracleous \& Forster 1998) and
calling for a substantial revision of AGN synthesis models for the
XRB (Gilli, Risaliti \& Salvati 1999). \\
 
An alternative possibility (see Comastri 2000) is that X--ray obscured AGN 
show a large variety of optical properties including 
those of SAXJ~1353.9$+$1820. It 
is worth noting that column densities as high as 10$^{23.5}$ cm$^{-2}$  
have been detected in Broad Absorption Line QSO (Gallagher et al. 1999), 
in some UV bright soft X--ray weak QSO (Brandt et al. 1999), 
and in a few HELLAS sources optically identified with broad line blue quasars (Fiore et al. 1999).
It is thus possible that the sources responsible for a 
large fraction of the XRB energy density are characterized by 
a large spread in their optical to X--ray properties. 
The $\alpha_{ox}$ spectral index, defined as the slope joining 
the 2500 $\AA$ and the 2 keV flux densities, is usually employed to measure the optical to X--ray ratio. 
Not surprisingly, absorbed objects are characterized 
by values of $\alpha_{ox}$ ($>$ 1.8) much steeper than the 
average value of bright unabsorbed quasars and Seyfert galaxies, 
$\alpha_{ox}$ $\simeq$ 1.5 (Laor et al. 1997; Yuan et al. 1998), 
while the faint nuclear UV flux density and the relatively bright 2 keV flux 
of SAXJ~1353.9$+$1820 correspond to $\alpha_{ox}$ $\simeq$ 1, which is quite 
flat but not unusual for red quasars (Kim \& Elvis 1999).   

A detailed discussion on the quasars contribution 
to the XRB is beyond the purposes of this Letter. 
Here we note that as far as the X--ray spectral properties 
and the bolometric luminosity of $\simeq$ 10$^{45}$ erg s$^{-1}$ 
(estimated using the average SED of Elvis et al. 1994 
with the measured $\alpha_{ox}$) are concerned, SAXJ~1353$+$1820 
can be classified as a high--luminosity absorbed AGN.
  
Future sensitive X--ray observations with Chandra and XMM coupled with 
optical spectropolarimetry data would be extremely helpful to 
better understand the nature of red quasars and to estimate 
their contribution to the XRB.

\section*{Acknowledgments}
We thank the {\it ASCA} team, who operate the satellite and maintain the software and 
database. We are also grateful to {\it BeppoSAX} Science Data Center and to the staff at the 
National Telescope Galileo who made possible our observations. 
The data discussed in this paper have been obtained within the 
TNG experimental phase programme (PI: F. La Franca \& R. Maiolino).
We thank G. Matt and L. Pozzetti for useful comments and an anonymous
referee for constructive suggestions. 
Financial support from Italian Space Agency under the contract ASI--ARS--98--119 
is acknowledged by C.~V. and A.~C. 
This work was partly supported by the Italian Ministry 
for University and Research (MURST) under grant Cofin98-02-32. \\


\begin{thebibliography}{99}

\bibitem{} Anders E., Grevesse N., 1989, Geochimica et Cosmochimica Acta, 53, 197
\bibitem{} Arnaud K.~A., in Jacoby G., Barnes J., eds, Astronomical Data Analysis Software 
and Systems V, Vol.~101, 17, ASP Conf. Ser., San Francisco
\bibitem{} Avni Y., 1976, ApJ, 210, 642
\bibitem{} Balucinska-Church M., McCammon D., 1992, ApJ, 400, 699
\bibitem{} Benn C.~S., Vigotti M., Carballo R., Gonzalez-Serrano J.~I., Sanchez S.~F., 1998, MNRAS, 295, 451
\bibitem{} Bohlin R.~C., Savage B.~D., Drake J.~F., 1978, ApJ, 224, 291
\bibitem{} Brandt W.~N., Fabian A.~C., Takahashi K., Fujimoto R., Yamashita A., 
Inoue H., Ogasaka Y., 1997a, MNRAS, 290, 617
\bibitem{} Brandt W.~N., Mathur S., Reynolds C.~S., Elvis M., 1997b, MNRAS, 292, 407
\bibitem{} Brandt W.~N., Laor A., Wills B.~J., 1999, ApJ, in press (astroph/9908016)
\bibitem{} Bruzual A.~G., Charlot S., 1993, ApJ, 405, 538
\bibitem{} Comastri A., Setti G., Zamorani G., Hasinger G., 1995, A\&A, 296, 1
\bibitem{} Comastri A., 2000, Astroph. Lett. \& Comm., submitted
\bibitem{} Dickey J.~M., Lockman F.~J., 1990, ARA\&A, 28, 215
\bibitem{} Elvis M., et al., 1994, ApJS, 95, 1
\bibitem{} Fiore F., La Franca F., Giommi P., Elvis M., Matt G., 
Comastri A., Molendi S., Gioia I., 1999, MNRAS, 306, L55
\bibitem{} Francis P.~J., Hewett P.~C., Foltz C.~B., Chaffee F.~H., 
Weymann R.~J., Morris S.~L., 1991, ApJ, 373, 465
\bibitem{} Fukugita M., Shimasaku K., Ichikawa T., 1995, PASP, 107, 945
\bibitem{} Gallagher S.~C., Brandt W.~N., Sambruna R.~M., Mathur S., Yamasaki N., 1999, ApJ, 519, 549
\bibitem{} Gendreau K., 1995, Ph.D. Thesis, Massachussets Inst. Tech.
\bibitem{} George I.~M., Turner T.~J., Yaqoob T., et al., 1999, ApJ, in press (astroph/9910218)
\bibitem{} Gilli R., Risaliti G., Salvati M., 1999, A\&A, 347, 424
\bibitem{} Halpern J.~P., Eracleous M., Forster K., 1998, ApJ, 501, 103
\bibitem{} Hayashi I., Koyama K., Awaki H., Ueno S., Yamauchi S., 1996, PASJ, 48, 219
\bibitem{} Kim D.-W., Elvis M., 1999, ApJ, 516, 9
\bibitem{} Laor A., Fiore F., Elvis M., Wilkes B.~J., McDowell J.~C., 1997, ApJ, 477, 93 
\bibitem{} Madau P., Ghisellini G., Fabian A.~C., 1994, MNRAS, 270, L17 
\bibitem{} Makishima K., et al., 1996, PASJ, 48, 171
\bibitem{} Masci F.~J., Webster R.~L., Francis P.~J., 1998, MNRAS, 301, 975
\bibitem{} Nandra K., Pounds K.~A., 1994, MNRAS, 268, 405
\bibitem{} Otani C., Dotani T., 1994, ASCA Newslett., 2, 25
\bibitem{} Reeves J.~N., Turner M.~J.~L., Ohashi T., Kii T., 1997, MNRAS, 292, 468
\bibitem{} Smith H.~E., Spinrad H., 1980, ApJ, 236, 419
\bibitem{} Tanaka Y., Inoue H., Holt S.~S., 1994, PASJ, 46, L37
\bibitem{} Vignali C., Comastri A., Stirpe G.~M., Cappi M., Palumbo G.~G.~C., Matsuoka M., 
Malaguti G., Bassani L., 1998, A\&A, 333, 411
\bibitem{} Webster R.~L., Francis P.~J., Peterson B.~A., Drinkwater M.~J., 
Masci F.~J., 1995, Nature, 375, 469
\bibitem{} Yuan W., Brinkmann W., Siebert J., Voges W., 1998, A\&A, 330, 108 
\end{thebibliography}
\end{document}